\documentclass[]{pasj01}
\draft

\begin{document} 
\Received{}
\Accepted{}

\title{Starspot activity and superflares on solar-type stars}

\author{Hiroyuki \textsc{Maehara}\altaffilmark{1}}%
\altaffiltext{1}{Okayama Astrophysical Observatory, 
National Astronomical Observatory of Japan, NINS}
\email{h.maehara@oao.nao.ac.jp}

\author{Yuta \textsc{Notsu},\altaffilmark{2}}
\altaffiltext{2}{Department of Astronomy, Graduate School of Science, Kyoto University}

\author{Shota \textsc{Notsu}\altaffilmark{2}}

\author{Kosuke \textsc{Namekata}\altaffilmark{2}}

\author{Satoshi \textsc{Honda}\altaffilmark{3}}
\altaffiltext{3}{Nishi-Harima Astronomical Observatory, Center for Astronomy, University of Hyogo}

\author{Takako T. \textsc{Ishii}\altaffilmark{4}}
\altaffiltext{4}{Kwasan and Hida Observatories, Graduate School of Science, Kyoto University}

\author{Daisaku \textsc{Nogami}\altaffilmark{2}}

\author{Kazunari \textsc{Shibata}\altaffilmark{4}}


\KeyWords{stars: activity --- stars: flare --- stars: solar-type --- starspots } 

\maketitle

\begin{abstract}
We analyze the correlation between starspots and superflares on solar-type
stars using observations from the Kepler mission.
The analysis shows that the observed fraction of stars with superflares
decreases as the rotation period increases and as the amplitude of photometric variability 
associated with rotation decreases.
We found that the fraction of stars with superflares among the
stars showing large-amplitude rotational variations, 
which are thought to be the signature of the large starspots,
also decreases as the rotation period increases.
The small fraction of superflare stars among the stars with large
starspots in the longer-period regime suggests that some of
the stars with large starspots show a much lower flare activity
than the superflare stars with the same spot area.
Assuming simple relations between spot area and lifetime and between spot temperature
and photospheric temperature, 
we compared the size distribution of large starspot groups on 
slowly-rotating solar-type stars with that of sunspot groups. 
The size distribution of starspots shows the power-law distribution 
and the size distribution of larger sunspots lies on this power-law line. 
We also found that frequency-energy distributions
for flares originating from spots with different sizes are the same 
for solar-type stars with superflares and the Sun.
These results suggest that the magnetic activity we observe on solar-type stars 
with superflares and that on the Sun is caused by the same physical processes.
\end{abstract}

\section{Introduction}
Solar flares are sudden energy releases in the solar atmosphere
caused by the magnetic reconnection (e.g., \cite{Shibata2011}).
The typical energy released by solar flares ranges from
$10^{28}$ to $10^{32}$ erg.
Similar flares and larger ``superflares'' have been observed on
various types of stars including solar-type stars (G- and 
late F-type main sequence stars; e.g., \cite{Weaver1973,Schaefer2000}). 
The recent space-based observations revealed that not only rapidly
rotating and young solar-type stars but also
slowly-rotating stars like our Sun exhibit superflares
with the bolometric energy of
$10^{33}$ to $10^{35}$ erg, which are $10$-$1000$ times larger than
the energy of the largest solar flares.
(e.g., \cite{Maehara2012, Shibayama2013, Maehara2015}).

Most of superflare stars show quasi-periodic light variations
with the period and amplitude of $\sim 1$-$30$ days and $\sim 0.1-10$ \%.
By analogy with the time variations in the total solar irradiance due
to sunspots (e.g., \cite{Kopp2005}),
the brightness variations of superflare stars are thought to be caused by 
the rotation of stars and the superflare stars have much larger starspots
than our Sun (e.g., \cite{Notsu2013}).
According to \citet{Shibata2013}, 
the energy released by superflares are basically 
a part of magnetic energy stored around starspots
if we assumed that the magnetic energy can be estimated
from the area of starspots 
and the starspot area corresponds to
the amplitude of light variations.

High-dispersion spectroscopy of superflare stars (\cite{Notsu2015a,Notsu2015b})
revealed that projected rotation velocity  $v\sin i$ of superflare stars
is consistent with the rotation velocity estimated from the period of
light variations.
\citet{Notsu2015b} found that there is a clear
correlation between the intensity of the chromospheric Ca II 8542 \AA\ 
 line and the amplitude of quasi-periodic light variations of superflare stars.
These results suggest that the photometric variations of superflare stars
are caused by the rotation of the star and the amplitude of brightness 
variations is a good indicator of the chromospheric activity.
Moreover, \citet{Karoff2016} reported that superflare stars show higher
intensities of the Ca II H and K lines than ordinary solar-type stars.
This result indicates that superflare stars show higher magnetic
activity 
than ordinary solar-type stars including our Sun.
Since the existence of large starspots (up to $\sim 10$ \% of
stellar hemisphere) is thought to be a key factor to produce superflares,
it is important to investigate the statistical properties of large starspots
(e.g., production rate and size distribution of starspots)
and their dependence on stellar properties
(e.g., rotation period) for the understanding of superflares.
In this paper, we report (1) the relation between
the flare activity and properties of starspots on solar-type stars,
(2) the production rate and size distribution of starspots on solar-type stars,
and (3)
the comparison between the starspot-superflare connection on solar-type stars and
the sunspot-solar flare connection on the Sun.

\section{Data and method}
\subsection{Sample selection}
We selected solar-type stars (early G- and late F-type main sequence stars)
from the Kepler data set
by using the effective temperature ($T_{\rm eff}$) and the surface
gravity ($\log g$) of the star. 
The dynamo activity is thought to correlate with the 
stellar mass, which is related to the effective temperature,
and rotation period of the star (e.g., \cite{Kippenhahn1990}).
In order to compare the spot
activity of the solar-type stars and that of the Sun, we select the 
solar-type stars with the temperature similar to the Sun.
Here 
we used the selection criteria of $5600 {\rm K} < T_{\rm eff} < 6300 {\rm K}$
and $\log g > 4.0$ for the selection of solar-type stars.
We also selected the stars exhibiting superflares from the flare star list
in \citet{Shibayama2013}.
In previous studies about superflares \citep{Maehara2012, Shibayama2013, Candelaresi2014}, 
we selected G-, K-, and M-dwarfs as the target of search for superflares by using stellar
parameters in the original Kepler Input Catalog \citep{Brown2011}. 
Since the temperature and surface gravity of the stars taken from \citet{Brown2011}
are slightly different from those in \citet{Huber2014} (e.g., systematic difference
in effective temperature),
some stars selected by using the above criteria 
are not included in the G-dwarf samples in \citet{Shibayama2013}.
In order to avoid mismatch of the data caused by 
the difference in stellar parameters between \citet{Brown2011} and \citet{Huber2014},
we selected samples that fulfill the both criteria:
(1) $5600 {\rm K} < T_{\rm eff} < 6300 {\rm K}$
and $\log g > 4.0$ in \citet{Huber2014};
(2)  $5100 {\rm K} < T_{\rm eff} < 6000 {\rm K}$
and $\log g > 4.0$ in \citet{Brown2011}, which is the same
criteria as those used
in \citet{Shibayama2013}.
The total number of samples is 64,239.
The typical length of observations to search for superflares is $\sim 480$ days \citep{Shibayama2013}.

\subsection{Bolometric energy of superflares}
We used the data set of superflares detected by \citet{Shibayama2013}.
However, since the effective temperature of the stars taken from 
\citet{Brown2011} is approximately 200 K higher than those from \citet{Huber2014}, 
we recalculated the bolometric energy of each superflare
by using the stellar parameters taken from \citet{Huber2014} and
the Kepler PDC light curve (DR24).
The bolometric energy emitted by each flare is estimated
by time integration of the change in the stellar luminosity
caused by the flare with the same manner as \citet{Shibayama2013}.

\subsection{Rotation period and amplitude of light variations}
\citet{McQuillan2014} derived the rotation period and median amplitude of
periodic variability of main-sequence stars with $T_{\rm eff}<6500$K.
In our analysis, rotation period and amplitude of light variations of 
our samples are taken from \citet{McQuillan2014}.
The rotation periods and amplitudes of 9,383 stars among 64,239 
solar-type samples were detected.
The typical observation length for searching for the period and amplitude of 
light variations
is $\sim 1400$ days \citep{McQuillan2014}.

\subsection{Area of starspots}
We estimated the area of starspots from the amplitude of brightness variations
and radius of the stars.
The normalized amplitude of rotational light variation of the star ($\Delta F/F$)
with the apparent starspot area ($A_{\rm spot}^{*}$) can be written as
\begin{equation}
\frac{\Delta F}{F} = \frac{A_{\rm spot}^{*}(T_{\rm star}^4 - T_{\rm spot}^{4})}{A_{\rm star}T_{\rm star}^4}, 
\label{obs_amplitude}
\end{equation}
where $A_{\rm star}$ indicates the apparent area of the star, $T_{\rm star}$ and $T_{\rm spot}$
are the temperature of unspotted photosphere of the star and starspot, respectively.
The temperature difference between stellar photosphere and starspots depends
on the photospheric temperature of the stars (e.g., \cite{Berdyugina2005}).
We estimated the temperature difference between photosphere and spots ($T_{\rm star} - T_{\rm spot}$)
by using the following equation,
\begin{equation}
\Delta T(T_{\rm star}) = T_{\rm star} - T_{\rm spot} = 3.58\times 10^{-5}T_{\rm star}^2  + 0.249T_{\rm star} - 808,
\label{deltaT}
\end{equation}
which was derived from the second order polynomial fit to the data for all stars except for EK Dra 
in table 5 of \citet{Berdyugina2005}.
The RMS of residuals of the fit is $220$ K.
Therefore, from equation (\ref{obs_amplitude}) and (\ref{deltaT}),
we can estimate the area of starspots normalized by the area of
solar hemisphere ($A_{\rm spot}$) as follows:
\begin{equation}
A_{\rm spot}=\left(\frac{R_{\rm star}}{R_{\odot}}\right)^{2} \frac{T_{\rm star}^4}{T_{\rm star}^4 - \{T_{\rm star} - \Delta T(T_{\rm star})\}^{4}} \frac{\Delta F}{F},
\label{spotarea}
\end{equation}
where $R_{\rm star}/R_{\odot}$ is stellar radius in units of the solar radius.
The equations (\ref{obs_amplitude}) and (\ref{spotarea}) are applicable only if
(a) there are a few large spot groups on the surface of the star
and (b) the lifetime of spots is much longer than the rotation period.
As discussed in appendix 1, in the case of the solar-type stars
showing large amplitude brightness variations,
the typical number of major spot groups which 
contribute to the photometric light variations is $1$-$3$.
The typical lifetime of such spots ranges from $\sim 50$ days to $\sim 300$ days,
which is longer than the rotation period of solar-type stars ($<40$ days).
Therefore we can estimate the area of large starspots on the
solar-type stars by using equation (\ref{spotarea}).
In the later analysis,
$R_{\rm star}/R_{\odot}$ and 
$T_{\rm star}$ are taken from \citet{Huber2014}
and $\Delta F/F$ are taken from \citet{McQuillan2014}.

\section{Results}
\begin{figure}
\begin{center}
\includegraphics{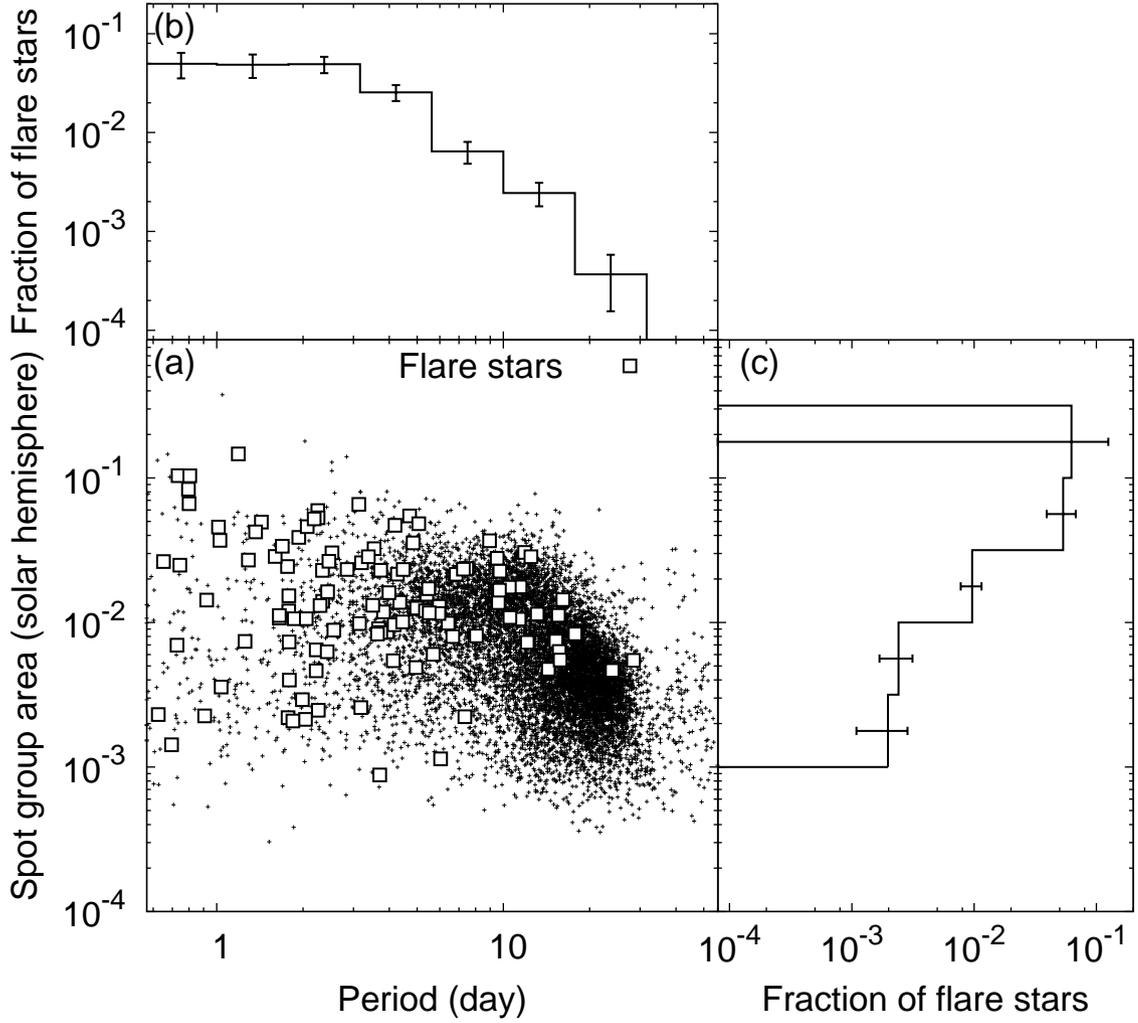}
\end{center}
\caption{(a) Scatter plot of the spot group area of solar-type stars as a 
function of the rotation period. The vertical axis represents the total 
area of starspots (in units of the area of solar hemisphere
 ($A_{1/2\odot} \sim 3.3\times 10^{22} \rm{cm}^{2}$) estimated from 
the amplitude of quasi-periodic light variations by using equation (\ref{spotarea}).
Open squares and small dots indicate solar-type stars with and without 
superflares respectively.
(b) Histogram of the fraction of the stars showing superflares as a
function of rotation period. The error bars represent the square root of 
the number of flare stars in each bin.
(c) Same as (b) but as a function of the spot group area.
}
\label{Aspot_vs_Prot}
\end{figure}

Figure \ref{Aspot_vs_Prot} (a) shows the scatter plot of the total spot-group
area of solar-type stars in units of the area of solar hemisphere 
($A_{1/2\odot} = 3\times 10^{22} \rm{cm}^{2}$)
as a function  of the rotation period ($P_{\rm rot}$).
Small dots indicate all solar-type stars ($5600<T_{\rm eff}<6300$K, 
$\log g > 4.0$) and open squares indicate solar-type stars showing
superflares.
Please note that there are many ``inactive'' stars whose area of starspots
$<10^{-3}\,A_{1/2\odot}$. Approximately 85 \% (54,856/64,239)
stars are not plotted in figure \ref{Aspot_vs_Prot} (a) 
since the amplitude of light variations is smaller than the detection limit.
The largest area of starspots on solar-type stars in a given period bin
is roughly constant ($\sim 5\times 10^{-2} A_{1/2\odot}$) in
the period range of $P_{\rm rot}<12$ days.
However, in the period range of $P_{\rm rot}>12$ days, the largest starspot
area decreases as the rotation period increases.
The stars showing superflares tend to have shorter rotation period
and larger starspot area.
Figure \ref{Aspot_vs_Prot} (b) and (c) show the observed fraction 
of the stars showing superflares as a function of the rotation 
period and area of starspots, respectively.
As shown in figure \ref{Aspot_vs_Prot} (b), the fraction of flare
stars decreases as the rotation period increases in the range of the 
rotation period $P_{\rm rot}>3$ days.
In the period range of $P_{\rm rot}<3$ days, the observed fraction of
superflare stars is almost constant.
The previous studies (e.g., \cite{Maehara2012}) pointed out that
the flare frequency as a function of the rotation period shows
the similar decrease trend and ``saturation'' in the period 
range of $P_{\rm rot}>3$ and $P_{\rm rot}<3$ days.
Moreover, the fraction of superflare stars also depends on the
total area of starspots.
Fraction of superflare stars rapidly decreases as the starspot area 
decreases as presented in figure \ref{Aspot_vs_Prot} (c).
In case of the stars whose area of starspots exceeds $10^{-1.5} A_{1/2\odot}$,
the fraction of superflare stars
is $\sim 8$\%. The fraction decreases to $\sim 0.8$\%, $\sim 0.5$\%, and
$\sim 0.02$\% as the spot area decreases to $10^{-2.0}$-$10^{-1.5} A_{1/2\odot}$, 
$10^{-2.5}$-$10^{-2.0} A_{1/2\odot}$, and $10^{-3.0}$-$10^{-2.5} A_{1/2\odot}$.

\subsection{Fraction of the stars with large starspots as a function of rotation period}
\begin{figure}
\begin{center}
\includegraphics{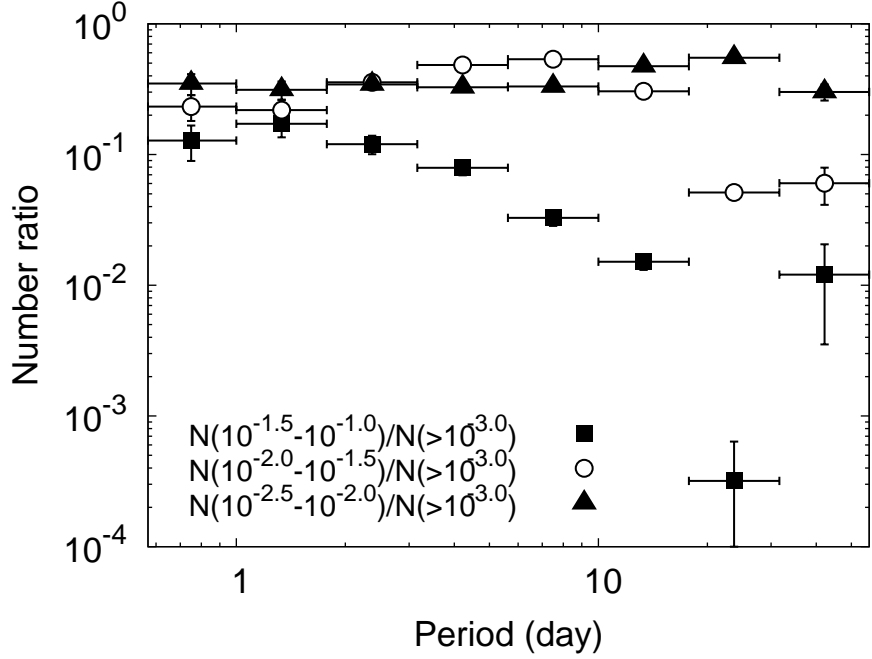}
\end{center}
\caption{
The fraction of the stars with large starspots among the stars
whose area of spots is larger than $10^{-3} A_{1/2\odot}$.
Filled squares, open circles, and filled triangles
 indicate the stars whose starspot area are
$10^{-1.5}$-$10^{-1.0}$,  $10^{-2.0}$-$10^{-1.5}$, and $10^{-2.5}$-$10^{-2.0} A_{1/2\odot}$, 
respectively.
The horizontal and vertical error bars represent the bin width
and the square root of 
the number of stars in each period bin respectively.
}
\label{Nstar_largespots}
\end{figure}

Figure \ref{Nstar_largespots} presents the ratio of the number of stars
with a given starspots area to the total number of stars with the starspot area
$> 10^{-3} A_{1/2\odot}$ as a function of the rotation period.
As mentioned above, 
the detection completeness for the
rotation period of the stars with the starspot area $< 10^{-3} A_{1/2\odot}$
is almost zero due to the detection limit.
Since it is difficult to estimate the correct total 
number of the stars in a given period bin,
here we used the number of the stars whose starspot area $>10^{-3} A_{1/2\odot}$
instead of the total number of stars in a given period bin.
The fraction of the stars with the starspot area of $10^{-1.5}$ - $10^{-1.0} A_{1/2\odot}$
(filled squares) decreases as the rotation period increases 
in the period range above $3$ days.
On the other hand, the fractions of the stars with the starspot area of 
$10^{-2.0}$ - $10^{-1.5} A_{1/2\odot}$ (open circles) and 
$10^{-2.5}$ - $10^{-2.0} A_{1/2\odot}$
(filled triangle) are almost constant 
in the ranges of the rotation period of $0.6 <P_{\rm rot} <  20$ days and $0.6 <P_{\rm rot} <  60$ days, respectively .

\subsection{Fraction of superflare stars as a function of rotation period}
\begin{figure}
\begin{center}
\includegraphics{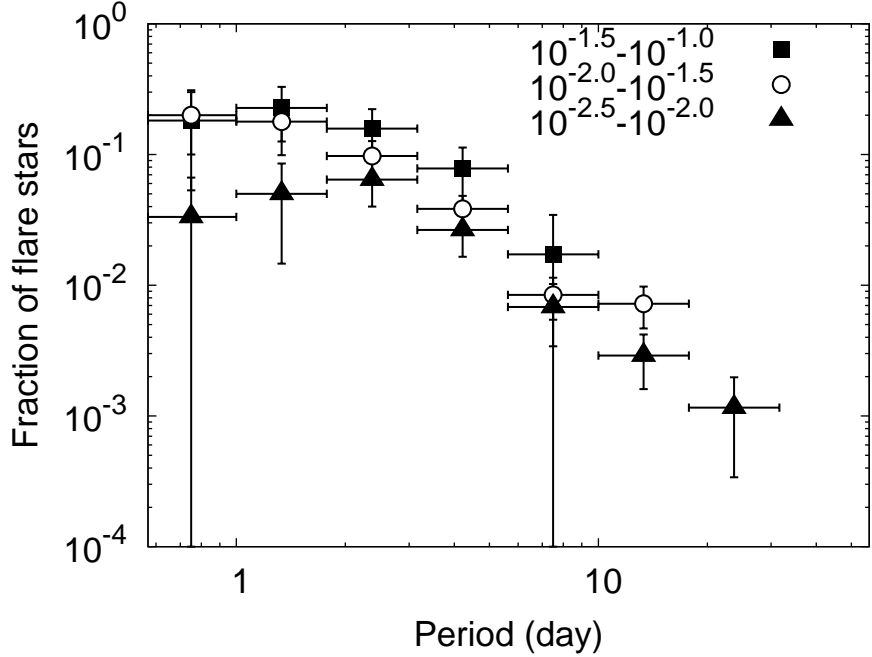}
\end{center}
\caption{
The fraction of superflare stars (stars showing superflares 
with $E_{\rm flare} > 10^{34}$ erg during the Kepler observations)
among the stars with a given spot area.
Filled squares, open circles and filled triangles indicate the flare star fraction among
the stars with the starspot area of 
$10^{-1.5}$-$10^{-1.0}$, $10^{-2.0}$-$10^{-1.5}$ and $10^{-2.5}$-$10^{-2.0} A_{1/2\odot}$
, respectively.
The horizontal and vertical error bars represent 
the bin width and 
the square root of 
the number of superflare stars in each period bin respectively.
}
\label{flare_star_fraction}
\end{figure}

Figure \ref{flare_star_fraction} shows the fraction of the stars showing
superflares with the bolometric energy of $>10^{34}$ erg
among the stars with a given starspot area
as a function of rotation period.
The fractions of superflare stars among the stars with the starspot area of
$10^{-1.5}$ - $10^{-1.0}$ $A_{1/2\odot}$ (filled squares), 
$10^{-2.0}$ - $10^{-1.5}$ $A_{1/2\odot}$ (open circles) 
and $10^{-2.5}$ - $10^{-2.0}$ $A_{1/2\odot}$ (filled triangles), 
are roughly constant in the period range of $< 3$ days.
However, the observed fraction of superflare stars decreases as the 
rotation period increases in the period range of $>3$ days.
This indicates that the large amount of the slowly-rotating stars 
with the moderately large starspots ($A_{\rm spot} \sim 10^{-2} A_{1/2\odot}$)
did not exhibit energetic superflares during the Kepler observations.


\section{Discussion}
\subsection{Frequency of large starspots on slowly-rotating stars}
\begin{figure}
\begin{center}
\includegraphics{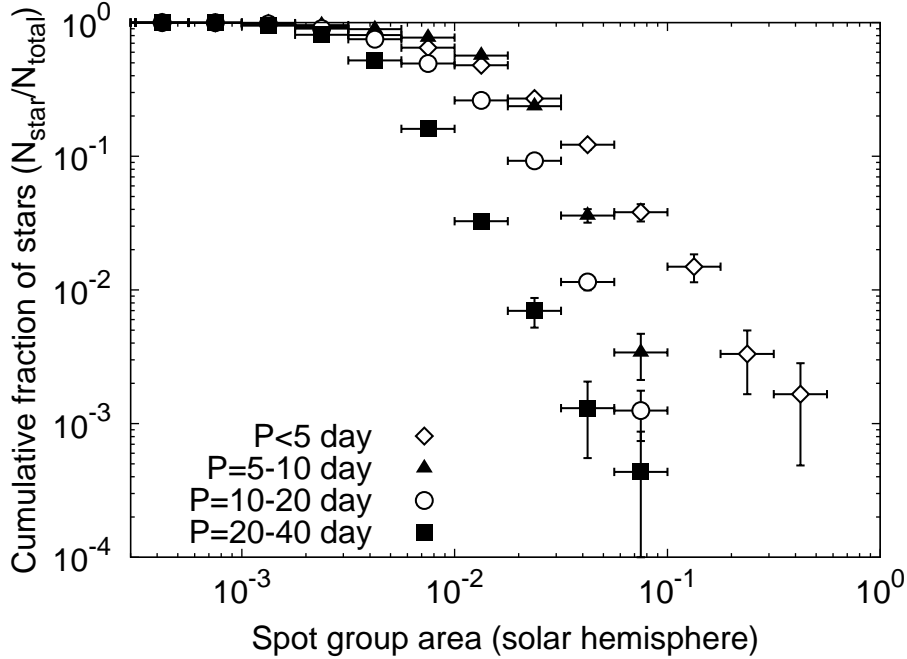}
\end{center}
\caption{
Cumulative fraction of stars as a function of the area of
starspots. The vertical axis indicates the number of stars
with the given spot area and rotation period range normalized by the
total number of stars in the given period range.
Open diamonds, filled triangles, open circles and filled squares
represent the cumulative fraction of solar-type stars with the rotation 
period of $<5$ days, $5$-$10$ days, $10$-$20$ days, and $20$-$40$ days.
The vertical and horizontal error bars indicate the square root of
the number of stars and bin width respectively.
}
\label{Nstar_vs_amp}
\end{figure}

Figure \ref{Nstar_vs_amp} shows the cumulative fraction
of the stars with a given spot area in each period range.
The cumulative fraction was calculated as the ratio of
the number of stars whose starspots area are
larger than or equal to a given value
to the total number of stars in each period range.
The cumulative number of stars becomes almost constant in the spot
area range below $10^{-3} A_{1/2\odot}$. 
As mentioned in the previous section, this caused by that the 
periodic light variations the stars with small starspots could
not be detected.
The fraction of the stars rapidly decreases as the 
area of starspots increases in the range of starspot area above
$10^{-2.5}$ - $10^{-2.0} A_{1/2\odot}$.
The fraction of the stars with a given starspot area increases as the 
rotation period decreases. These results indicate that the average
magnetic activity on rapidly-rotating stars is higher than
that on slowly-rotating stars.

If we assume that (1) the typical lifetime of
starspot groups is longer than the rotation period of the stars,
and (2) most of our samples have the similar stellar properties,
the number of stars with a give starspot area would be proportional to
the appearance frequency of starspots with a given area.
Under these assumptions, 
the number distribution of the stars
as a function of the starspot area would correspond
to the appearance frequency distribution of 
the starspots as a function of the starspot area.
According to \citet{Petrovay1997}, the lifetime of sunspots increases as the
area of sunspot increases. If the relation between lifetime of the spot
and spot area in the Sun is similar to that in the solar-type stars,
the lifetime of large starspots ($A>10^{-3} A_{1/2\odot}$) is longer than
the mean rotation period of the stars in this analysis ($\overline{P_{\rm rot}} \sim 30$ days).
As shown in appendix 1,
the solar-type stars with large amplitude ($\sim 1$ \%) 
photometric variations have large starspots with the
area of the order of $\sim 10^{-2} A_{1/2\odot}$ and 
the lifetime of such large spots ranges from 
$\sim 50$-$\sim 300$ days which 
is longer enough than the rotation period of the star.
We compared the frequency-area distribution
of starspots on slowly-rotating solar-type stars 
($P_{\rm rot}=20$-$40$ days) and that of sunspot groups.
As described in section 2.4, since the area of starspots were derived from
the median of the amplitude of light variations within one rotation period
taken from \citet{McQuillan2014},  we can estimate only the 
typical size of the largest starspot group on each star.
Let $N_{\rm star}(A)$ be a number of stars with the starspot area of 
$\geq A$. 
The fraction of the stars with the starspot area in excess
of $A$, $N_{\rm star}(A)/N_{\rm star, total}$, is equivalent to
the expected number of the largest starspots with the area of 
$\geq A$ per star during one rotation period.

As mentioned in the section 3.1, it is difficult to estimate the correct
number of $N_{\rm star,total}$ from the data set
because the period and amplitude of
rotational light variations of the stars with small starspots could not
be detected.
We estimated the total number of stars with the rotation period of 20-40 days
from the empirical gyrochronology relation (e.g., \cite{Barnes2007,
Mamajek2008, Meibom2009, Meibom2015}).
Let $N_{\rm star, total}(P)$ to be the
number of stars with the rotation period of $\geq P$ in the sample.
$N_{\rm star, total}(P)$ can be estimated from the duration of
the main sequence phase ($\tau _{\rm MS}$), the gyrochronological age
of the star ($t_{\rm gyro}(P)$), and 
the total number of the samples ($N_{\rm all}$)
 as follows:
\begin{equation}
N_{\rm star, total}(P) = \left( 1 - \frac{t_{\rm gyro}(P)}{\tau _{\rm MS}} \right) N_{\rm all},
\end{equation}
if we assume that the star formation
rate around the Kepler field has been roughly constant
for $\tau _{\rm MS}$.
According to \citet{Mamajek2008}, the age of the star with the temperature of 
$\sim 5800$K ($B-V \sim 0.65$) and rotation period of $>20$ days
is $>2.9\pm 0.5$ Gyr.
This suggests that $\sim 70\pm 10$ \% of the stars with the 
temperature of $\sim 5800$K have the rotation period
longer than 20 days, and $N_{\rm star, total}$ for solar-type stars
with $P_{\rm rot}=20$-$40$ days in our sample would be $\sim 45,000 \pm 5000$.
As mentioned above, 
the typical lifetime of the spots with the area of $> 10^{-3} A_{1/2\odot}$
is thought to be longer than the mean rotation period of the star ($\overline{P_{\rm rot}} \sim 30$ days).
This suggests that we can ignore the appearance and disappearance of starspots during one rotation period
which would affect the the appearance frequency of starspots.
Therefore the appearance frequency of starspot groups 
(the average number of starspot groups per star per unit time)
with the area of $\geq A$ averaged over one rotation period can be written as
\begin{equation}
\overline{N}_{\rm starspot}(A) = \frac{N_{\rm star}(A)}{\overline{P_{\rm rot}}N_{\rm star, total}}.
\end{equation}

The appearance frequency of sunspot groups with the maximum area of the sunspot group of $\geq A$ can be written as
\begin{equation}
\overline{N}_{\rm sunspot}(A) = \frac{N_{\rm sunspot}(A)}{T_{\rm obs, total}},
\end{equation}
where
$N_{\rm sunspot}(A)$ and $T_{\rm obs, total}$ are
the number of sunspot groups
with the maximum area of $\geq A$ and
the total time span of observations respectively.
We analyzed the Royal Observatory, Greenwich - USAF/NOAA Sunspot Data 
between 1874 and 2015\footnote[1]
{The data are retrieved from http://solarscience.msfc.nasa.gov/greenwch.shtml.}
and calculated the appearance frequency of sunspot groups
as a function of the sunspot group area.

\begin{figure}
\begin{center}
\includegraphics{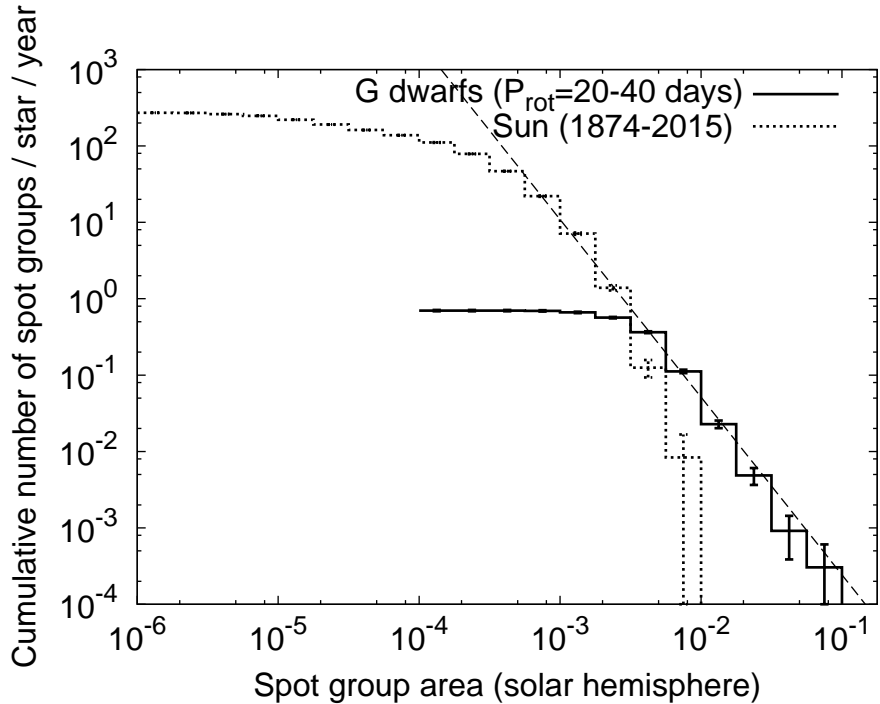}
\end{center}
\caption{
Comparison between the appearance frequency - spot area distribution of 
starspots on slowly-rotating solar-type stars and that of sunspot groups.
Solid lines indicate the cumulative appearance frequency of starspots
on the solar-type stars with the rotation period of 20-40 days.
Dotted lines indicate the cumulative appearance frequency of sunspot groups
as a function of the maximum area of each sunspot group.
The thin dashed line represents the power-law fit to the frequency distribution
of starspots in the spot area range of $10^{-2.5}$-$10^{-1.0} A_{1/2\odot}$.
The power-law index of the line is $-2.3\pm0.1$.
}
\label{comparison_sun_and_star}
\end{figure}

Figure \ref{comparison_sun_and_star} represents the comparison between
the cumulative appearance frequency ($\overline{N}(A)$) of starspots on 
solar-type stars with the rotation period of 20-40 days (solid lines)
and that of sunspot groups (dotted lines).
The occurrence frequency of decreases as the 
area of sunspots increases. 
The both appearance rate of the sunspot groups and that of starspots
with the area $>10^{-2.5} A_{1/2\odot}$
is approximately once in a few years.
The cumulative appearance frequency ($\overline{N}(A)$) of starspots on
solar-type stars can be fitted by a power-law function with the power-law
index of $-2.3\pm 0.1$ (dashed line) for the spot area between $10^{-2.5}$ to $10^{-1.0}$.
According to \citet{Bogdan1988}, the size distribution of individual sunspot umbral areas
shows the lognormal distribution. Although, the overall cumulative appearance frequency-size
distribution of sunspot groups also shows the similar lognormal distribution,
the size distribution of sunspot groups for large sunspots is roughly on this
power-law line for the spot area between $10^{-3.5} A_{1/2\odot}$ to $10^{-2.5} A_{1/2\odot}$.
The appearance frequency of sunspots with the spot area  of $\sim 10^{-2} A_{1/2\odot}$
is about 10 times lower than that of starspots on solar-type stars.
This difference between the Sun and solar-type stars might be caused by the lack
of ``super-active'' phase on our Sun during the last 140 years (e.g., \cite{Schrijver2012}).
The similarity between the size distribution of sunspots and that of starspots
implies that the both sunspots and larger starspots might be 
produced by the same physical process.


\subsection{Frequency of superflares and starspot area}
\begin{figure}
\begin{center}
\includegraphics{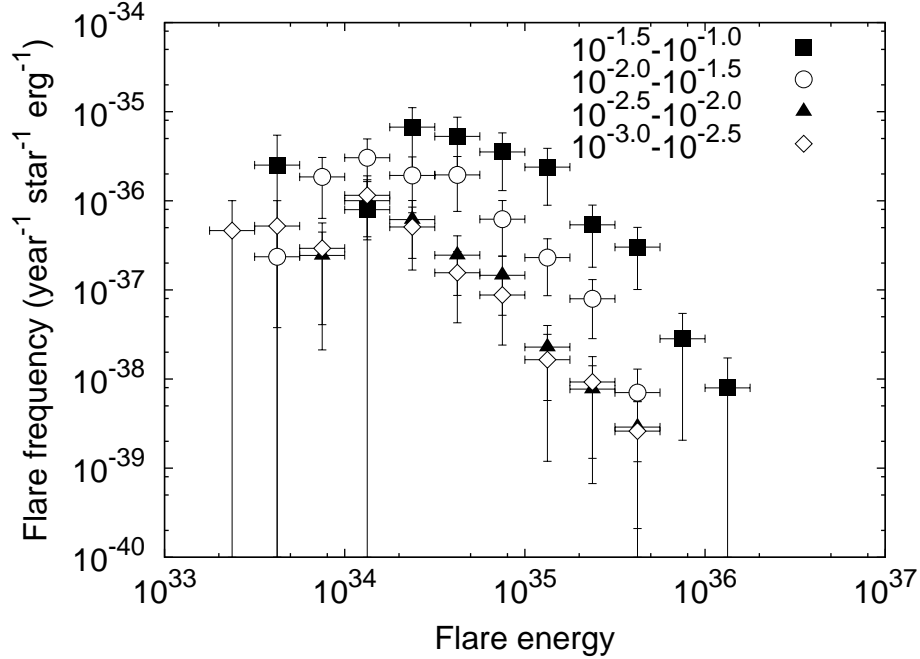}
\end{center}
\caption{
Occurrence frequency distribution of superflares
on solar-type stars with different starspot area.
Filled squares, open circles, filled triangles and open diamonds
represent the flare frequency on solar-type stars with 
starspot area of $10^{-1.5}$-$10^{-1.0}$, $10^{-2.0}$-$10^{-1.5}$,
$10^{-2.5}$-$10^{-2.0}$, and $10^{-3.0}$-$10^{-2.5}  A_{1/2\odot}$, respectively.
The vertical error bars indicate the square-root of the number of detected
flares. The horizontal error bars indicate the bin width.
}
\label{superflare_starspot_area}
\end{figure}

As shown in figure \ref{flare_star_fraction}, 
the fraction of the stars showing superflares
increases as the area of starspots increases.
This suggests that the average occurrence frequency of superflares
increases as the starspot area increases.
In addition to the fraction of flare stars and flare frequency,
the energy released
by the largest superflare also increases as the
area of starspots increases (e.g., \cite{Shibata2013, Maehara2015}).
Figure \ref{superflare_starspot_area} represents the 
frequency distribution of 
superflares on solar-type stars with different starspot area
as a function of the energy released by flares.
The frequency of superflares on solar-type stars with a given
starspot area is calculated from the number of solar-type stars
with a given starspot area, the total time span of observations,
and the number of flares which occurred on the solar-type stars
with a given starspot area.

The frequency of superflares with a given flare energy
clearly increases as the area of starspots increases.
The larger end of the frequency distribution is also increases
as the starspot area increases.
The larger end of the frequency distribution of superflares
on solar-type stars with the spot area of $\sim 10^{-2.5} A_{1/2\odot}$ is
$\sim 3\times 10^{35}$ erg and that on the stars with the spot
area of $\sim 10^{-1.5} A_{1/2\odot}$ is $\sim 10^{36}$ erg.

\begin{figure}
\begin{center}
\includegraphics{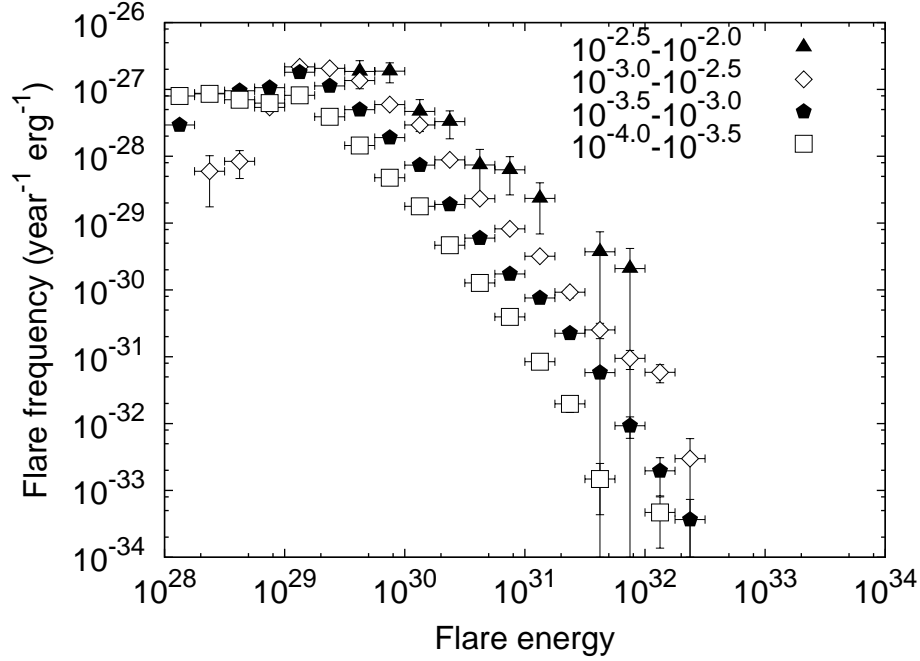}
\end{center}
\caption{
Occurrence frequency distribution of solar flares on active regions (ARs)
 with different sunspot group area.
Filled triangles, open diamonds, filled pentagons, and open squares
represent the flare frequency on ARs with sunspot group area
of $10^{-2.5}$-$10^{-2.0}$, $10^{-3.0}$-$10^{-2.5}$, $10^{-3.5}$-$10^{-3.0}$,
and $10^{-4.0}$-$10^{-3.5} A_{1/2\odot}$, respectively.
}
\label{solar_flare_sunspot_area}
\end{figure}

We adopted the same analysis to solar flares.
We used the event list of solar flares observed with
the GOES satellite from 1976 to 2015\footnote[2]
{The data are retrieved from ftp://ftp.ngdc.noaa.gov/STP/space-weather/solar-data/solar-features/solar-flares/x-rays/goes/xrs/.}.
Combined with the list
of solar flares and the  Royal Observatory, Greenwich - USAF/NOAA
Sunspot Data, we calculated the frequency of solar flares on
active regions (ARs) with a given sunspot area
from the number of flares on ARs, the number $\times$ duration of
ARs with a given area.
We assumed that bolometric energies of B, C, M, X, and X10 class 
solar flares  are $10^{28}$, $10^{29}$, $10^{30}$, $10^{31}$, 
and $10^{32}$ erg
from observational estimates of energies of typical solar flares
(e.g., \cite{Benz2008, Emslie2012}).
Figure \ref{solar_flare_sunspot_area} shows the frequency distribution of 
solar flares on ARs with different sunspot group area.
The occurrence frequency of flares increases as the sunspot
group area increases.

\begin{figure}
\begin{center}
\includegraphics{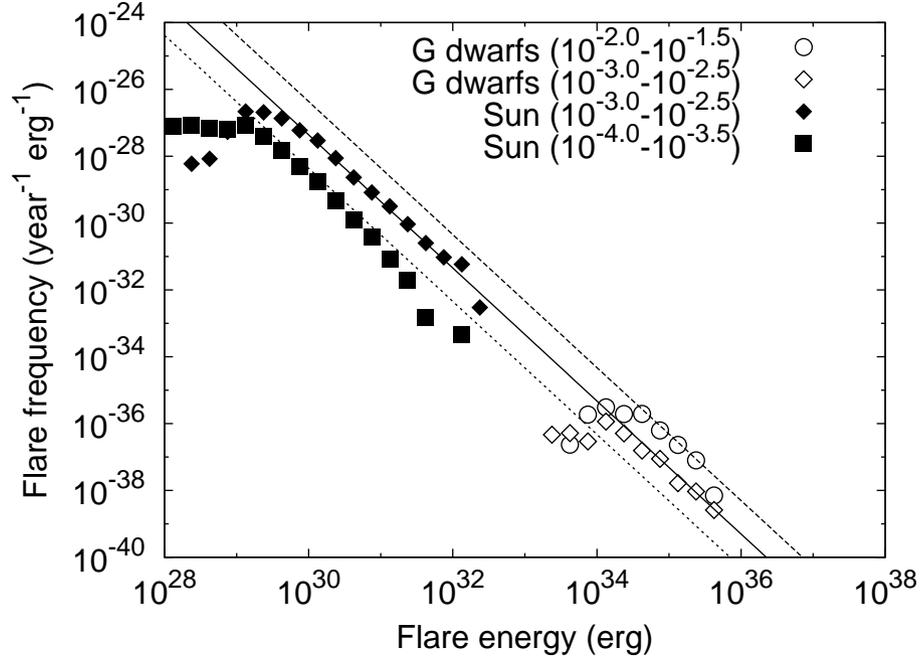}
\end{center}
\caption{
Comparison between occurrence frequency distribution of superflares 
on solar-type stars and that of solar flares.
Open circles and open diamonds represent the frequency of superflares
on solar-type stars with spot area of $10^{-2.0}$-$10^{-1.5}$
and $10^{-3.0}$-$10^{-2.5} A_{1/2\odot}$.
Filled diamonds and filled squares represent the frequency of solar
flares on active regions (ARs) with the sunspot group area of
$10^{-3.0}$-$10^{-2.5}$ and $10^{-4.0}$-$10^{-3.5} A_{1/2\odot}$.
Solid line indicates the power-law fit to the frequency of solar flares on
ARs with sunspot group area of $10^{-3.0}$-$10^{-2.5} A_{1/2\odot}$
between $10^{30}$ erg and $10^{33}$ erg. The power-law index is $-1.99\pm 0.05$.
Dashed and dotted lines are the power-law distributions with the same
power-law index but $10$- and $1/10$- times flare frequency of 
the solid line.
}
\label{solarflare_superflare_comp}
\end{figure}

Figure \ref{solarflare_superflare_comp} shows the comparison between
the occurrence frequency distribution of superflares on solar-type
stars (open circles and open diamonds) and that of solar flares
(filled diamonds and filled squares).
The solid line indicates the power-law fit to the 
frequency distribution of solar flares on ARs with sunspot group 
area of $10^{-3.0}$-$10^{-2.5} A_{1/2\odot}$
between $10^{30}$ erg and $10^{33}$ erg.
The power-law index is $-1.99\pm 0.05$.
Both the occurrence frequency distribution of superflares on solar-type
stars with the starspot area of $10^{-3.0}$-$10^{-2.5} A_{1/2\odot}$ (open diamonds)
and that of solar flares on ARs with the same sunspot group area 
($10^{-3.0}$-$10^{-2.5} A_{1/2\odot}$)
are roughly on the same power-law line.
Dashed and dotted lines in figure \ref{solarflare_superflare_comp} 
indicate power-law distribution with the same power-law index
but 10 and 1/10 times frequency of the solid line.
The frequency distribution of solar flares on ARs with sunspot group
area of $10^{-4.0}$-$10^{-3.5} A_{1/2\odot}$
and that of superflares on solar-type
stars with the starspot area of $10^{-2.0}$-$10^{-1.5} A_{1/2\odot}$
are roughly on the dotted and dashed lines respectively.
These results suggest that both the frequency of flares is roughly
proportional to the area of spot groups.

\subsection{Flare activity and  magnetic structure of spot groups}
\begin{figure}
\begin{center}
\includegraphics{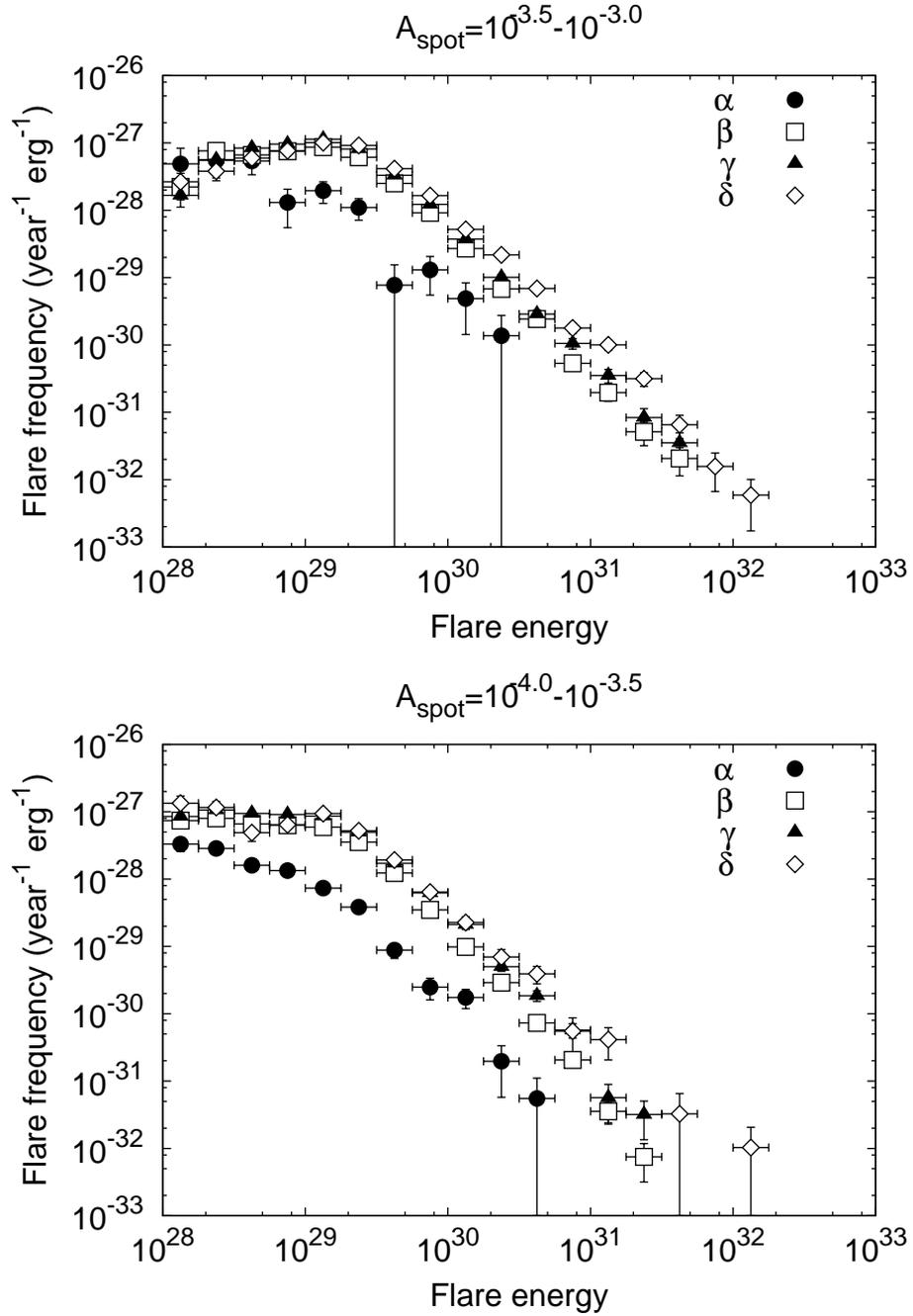}
\end{center}
\caption{
Occurrence frequency distribution of solar flares on active regions (ARs)
with 
different magnetic classifications.
Upper and lower panels show frequency distribution of solar flares 
on active regions (ARs) with the sunspot group area of
$10^{-4.0}$-$10^{-3.5}$ and $10^{-3.5}$-$10^{-3.0} A_{1/2\odot}$.
Filled circles, open squares, filled triangle, and open diamonds
represent the flare frequency on $\alpha$, $\beta$, $\gamma (\gamma + \beta\gamma)$,
and $\delta (\beta\delta + \gamma\delta + \beta\gamma\delta)$-type sunspot groups,
respectively.
The horizontal error bars indicate the bin width and the vertical error bars correspond to
the square root of the number of flares.
}
\label{solar_flare_sunspot_magnetic_class}
\end{figure}

The flare activity correlates with not only the area of sunspot 
groups, but also magnetic structure of sunspot groups
(e.g., \cite{Bell1959, Sammis2000}).
We performed the same analysis as in figure \ref{solar_flare_sunspot_area}
for solar flares on different magnetic classifications.
The magnetic classifications of each sunspot group were taken from
the USAF/NOAA Sunspot Data observed between 1981 and 2015\footnote[3]
{The data are retrieved from ftp://ftp.ngdc.noaa.gov/STP/SOLAR\_DATA/SUNSPOT\_REGIONS/USAF\_MWL/.}.
Figure \ref{solar_flare_sunspot_magnetic_class} represents
the frequency distribution of solar flares on ARs with the same
sunspot area but different magnetic structures.
As previous studies show, the occurrence frequency of flares 
increases as the magnetic structure of the sunspot groups becomes
complicated. 
The frequency of flares with a given flare energy on
the bipolar ($\beta$-, $\gamma$-, and $\delta$-type) sunspot groups 
is $1$-$2$ orders of magnitude
 higher than 
that on the unipolar ($\alpha$-type) sunspot groups. 
$\delta$-type sunspot groups exhibit more frequent 
flares than $\beta$- and $\gamma$-type sunspot groups.
The magnetic structure of sunspot groups also 
has a relation to the productivity of large solar flares.
According to \citet{Sammis2000}, more than 80\% of
X-class flares occurred on $\delta$-type spot groups.
As shown in figure \ref{solar_flare_sunspot_magnetic_class},
the largest end of the
frequency-energy distribution on $\delta$-type sunspot groups 
is also larger than those on other types of sunspot groups
with the same spot area.

\begin{figure}
\begin{center}
\includegraphics{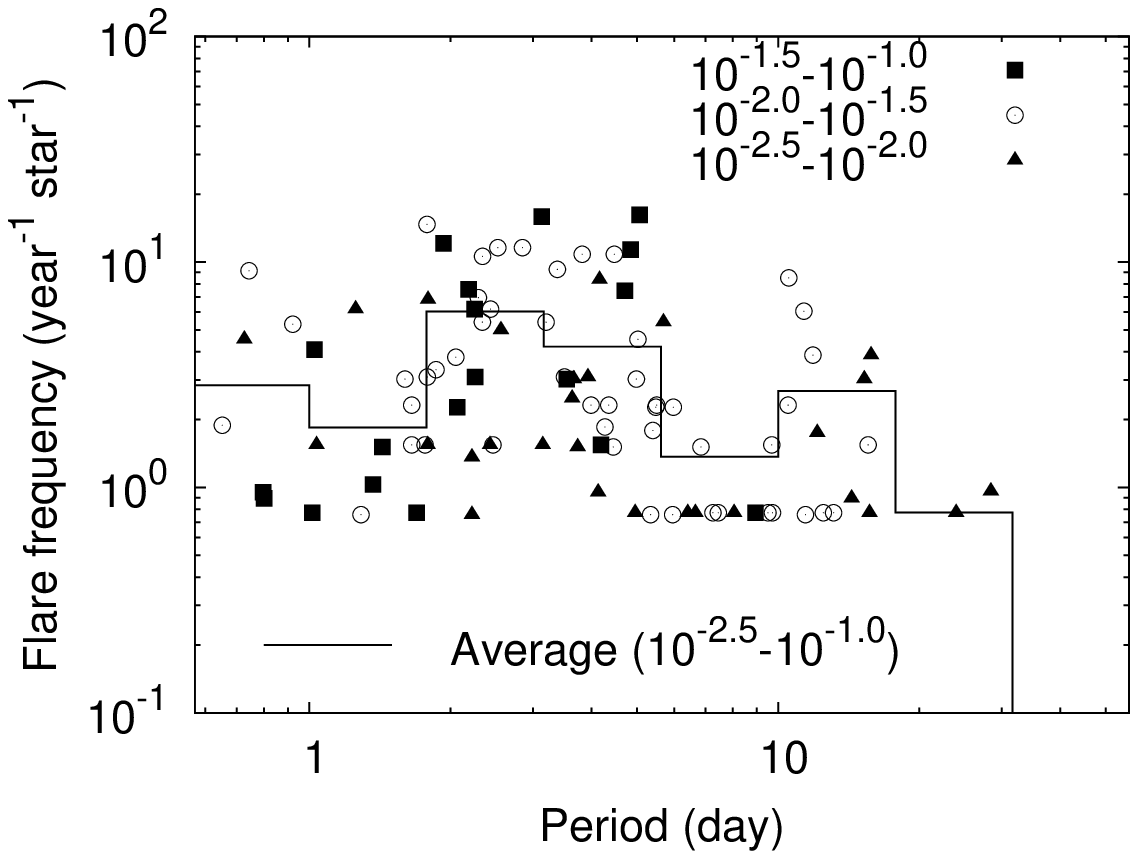}
\end{center}
\caption{
The frequency of superflare with $E_{\rm flare} > 10^{34}$ erg on 
the superflare stars as a function of the rotation period.
Filled squares, open circles, and filled triangles indicate the frequency
of superflares ($E_{\rm flare} > 10^{34}$ erg) on each superflare
star with the area of starspots of $10^{-1.5}$-$10^{-1.0}$, $10^{-2.0}$-$10^{-1.5}$,
and $10^{-2.5}$-$10^{-2.0} A_{1/2\odot}$, respectively.
The solid line histogram represents the average frequency of 
superflares on superflare stars with the area of starspots $>10^{-2.5} A_{1/2\odot}$.
Since the typical length of observation period for the flare detection 
is $\sim 480$ days, the lower limit
of the flare frequency is $\sim 1 {\rm event}/480 {\rm days} \sim 0.76$ year$^{-1}$.
}
\label{superflare_freq_vs_prot}
\end{figure}

As mentioned in section 3.2, 
the observed fractions of the stars showing superflares with the 
energy $>10^{34}$ erg
among the stars with the starspot area of $10^{-1.5}$-$10^{-1.0}$,
$10^{-2.0}$-$10^{-1.5}$, and $10^{-2.5}$-$10^{-2.0}$ $A_{1/2\odot}$
decreases as the rotation period increases.
Since the total time span of the observations for 
the detection of superflares is limited ($\sim 480$ days), the 
observed fraction of superflare stars can be affected by the frequency of
superflares.
Figure \ref{superflare_freq_vs_prot} represents
the frequency of superflares on each superflare star as a function of the rotation period.
The average flare frequency on the rapidly-rotating superflare stars ($P_{\rm rot}=2$-$3$ days)
is a few times higher than that on the superflare stars with $P_{\rm rot}>6$ days.
The rate of decrease in the frequency of superflares as a function of rotation period is
smaller than that in the fraction of stars vs. rotation period.
It should be noted that \citet{Davenport2016} reported many flare stars with 
the rotation period below $2$-$3$ days and some of them
are not listed in \citet{Shibayama2013}. On the other hand,
only a few flare stars with $P_{\rm rot} > 3$ days are listed as a flare star by 
\citet{Davenport2016} as mentioned in appendix 2. This difference may be caused 
by the differences in the detection method and selection criteria.
Since the flare detection threshold used in \citet{Shibayama2013} depends on
the rotation period and some superflares on rapidly-rotating stars may be missed, 
the flare frequency on rapidly-rotating superflare stars may be underestimated.

The frequency-energy distributions of superflares on the stars with
different spots sizes (figure \ref{superflare_starspot_area} and \ref{solarflare_superflare_comp}) indicate that 
the average frequency of superflares with the energy of $E_{\rm flare}>10^{34}$ erg
on the stars with the starspot area of $10^{-2.0}$-$10^{-1.5} A_{1/2\odot}$
is $\sim 10^{-1}$ year$^{-1}$.
The observed frequency of each flare star with $P_{\rm rot}>6$ days is
$\sim 1$ year$^{-1}$ (figure \ref{superflare_freq_vs_prot}). Therefore, 
the fraction of flare stars expected from $\sim 500$ days observations
would be $\sim 10^{-1}$. However, as shown in figure \ref{flare_star_fraction},
the observed fraction of the stars with superflares among the stars
with the starspot area of $10^{-2.0}$-$10^{-1.5} A_{1/2\odot}$ and
rotation period of $>6$ days is $\sim 10^{-2}$. This suggests
that some of the stars with larger starspots and longer rotation period
 may show a much lower flare activity than the superflare stars with
the same spot size.
Moreover, the frequency of 
superflares on each star with the same spot size and rotation period
have a large scatter ($>1$ order of magnitude).
These results imply that not only 
the existence of large starspots as previous studies pointed out but also other 
factors may play important roles in generation of superflares.
By analogy with the correlation between the flare activity and 
the magnetic structure of sunspot groups,
one of the possible explanation for
the difference in flare activity on the stars
with the same spot area may be the difference in the
magnetic structure of starspots.
As shown in figure \ref{solar_flare_sunspot_magnetic_class}, 
both the flare frequency 
and the flare energy at the larger end of frequency-energy 
distribution of solar flares originated from the $\alpha$-type spots are 
$1$-$2$ orders of magnitude lower than those from the $\delta$-type 
spots with the same spot area. 
If the correlation between the magnetic structure of large starspots and
superflares on the solar-type stars 
is similar to that between the magnetic structure of smaller sunspots and
solar flares on the Sun, the large starspots with the simple 
($\alpha$-type) magnetic structure would produce less frequent
and less energetic flares than the large starspots with the 
complex ($\delta$-type) magnetic structure.
According to sunspot observations (e.g., \cite{Sammis2000}), 
majority of sunspot groups with large spot area
are $\delta$-type sunspot group.
The USAF/NOAA Sunspot Data indicate that the fraction of  $\delta$-type spot groups
among the sunspots with the area of $10^{-3.0}$-$10^{-2.5} A_{1/2\odot}$ is approximately
92\%. However, the small fraction of the stars with superflares among the stars with large
starspots and longer rotation period (figure \ref{flare_star_fraction}) suggests that 
if the difference in the flare activity of the stars with large starspots is
caused by the difference in the magnetic structure of spots,
the fraction of the $\delta$-type 
spots among the starspots with the area of $\sim 10^{-2.0} A_{1/2\odot}$ on the solar-type
stars 
would be lower than that among the sunspots with the area of $\sim 10^{-3.0} A_{1/2\odot}$.
We need more detailed studies of the magnetic 
structure of
large starspots on the stars with and without superflares.
Doppler imaging (e.g., \cite{Rice2002}) and Zeeman Doppler imaging (e.g., \cite{Donati2009}) techniques
may help us to study the magnetic field geometry of the solar-type stars with
large starspots and its correlation to the flare activity.
The large starspots can cause the 
small fluctuations in the light curve of planetary transits (e.g., \cite{Silva2003, Pont2007}).
The detailed analysis of the transit light curve with the future
high photometric-precision and high time-resolution data
(e.g., {\it TESS}; \cite{Ricker2015} and {\it PLATO}; \cite{Rauer2014})
may also be useful to reveal the difference in the structure
of the starspot
groups between the stars with and without superflares.

\begin{ack}
Kepler was selected as the tenth Discovery mission. 
Funding for this mission is provided by the NASA Science Mission Directorate. 
The data presented in this paper were obtained from the Multimission Archive at STScI. 
This work was supported by MEXT/JSPS KAKENHI Grant Number 26800096, 26400231, 16H03955,
16J00320, and 16J06887.
The authors wish to acknowledge the anonymous reviewer for his/her 
helpful comments to the manuscript.
\end{ack}

\appendix 
\section{Large starspots on solar-type stars}
As described in section 2, we estimated the area of starspot groups 
from the amplitude
of the rotational light variations under the assumption that
observed light variations are caused by one or a few large
starspot group(s).
In the case of rapidly-rotating stars, the existence of the large
starspots has been studied by using the light curve modeling
(e.g., \cite{Frohlich2012, Notsu2013}) and Doppler imaging technique (e.g., \cite{Strassmeier1998, Rice2002}).
However, in the case of slowly-rotating stars, 
it is unclear whether the large amplitude light-variations
are caused by a few large starspot groups or not.
In order to check the assumption that light variations are
caused by the large starspots,
here we investigate the number of starspot groups
on some of stars showing the light variations with large 
amplitude ($>1$ \%) and long period ($>20$ days).

\begin{figure}
\begin{center}
\includegraphics[width=12cm]{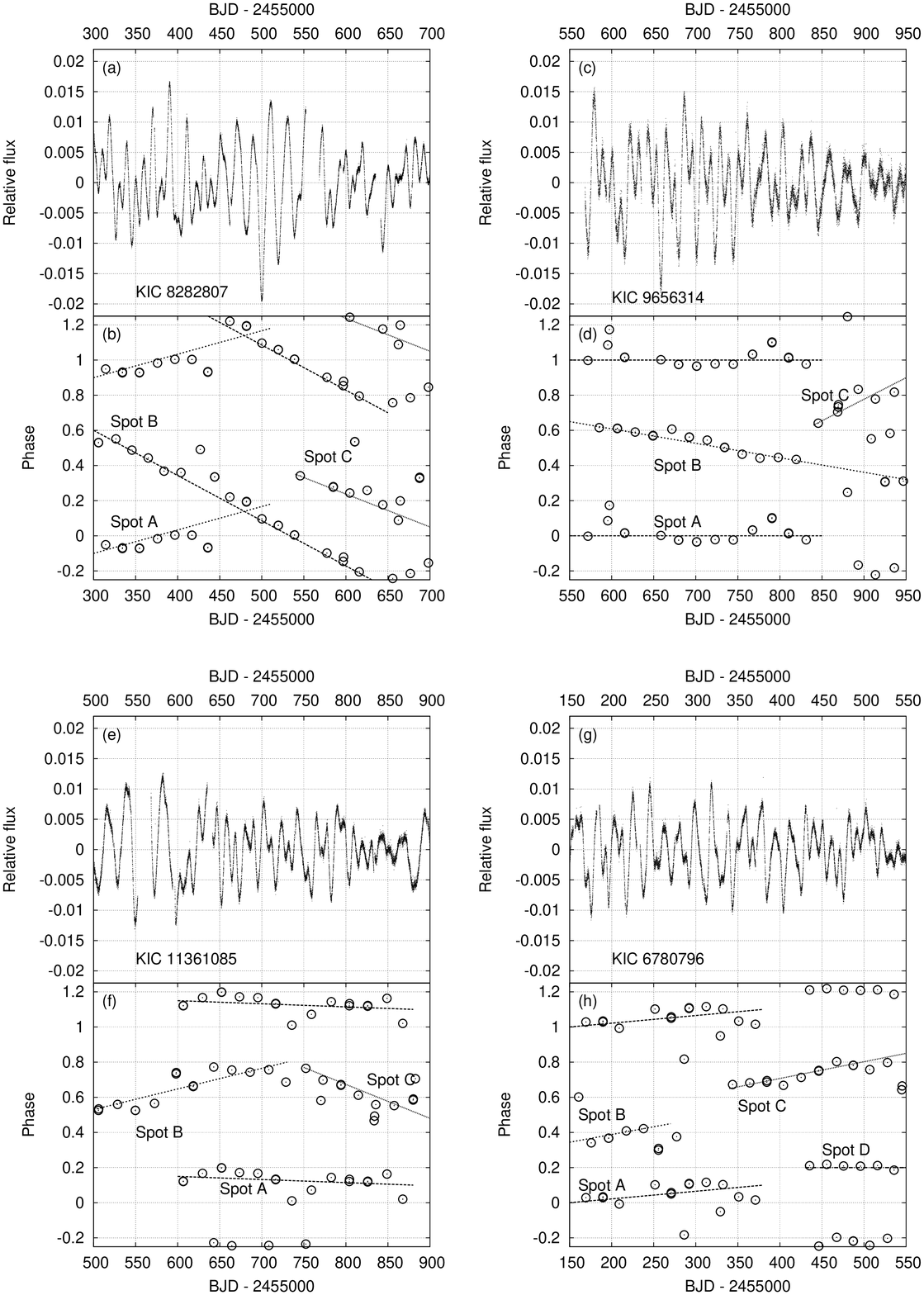}
\end{center}
\caption{
(a): Light curve of the slowly-rotating solar-type star KIC 8282807 ($P_{\rm rot}=20.3$ days).
(b): Time-variation of the rotational phase of the local photometric 
minima in the light curve of KIC 8282807. 
The rotational phase are calculated by using the rotation period taken from \citet{McQuillan2014}.
Open circles represent each local brightness minimum.
Solid and dotted lines indicate each series of local minima 
caused by each spot group.
(c) and (d): Same as panels (a) and (d), but for KIC 9656314 ($P_{\rm rot}=21.7$ days).
(e) and (f): Same as panels (a) and (d), but for KIC 11361085 ($P_{\rm rot}=21.9$ days).
(g) and (h): Same as panels (a) and (d), but for KIC 6780796 ($P_{\rm rot}=20.2$ days).
}
\label{spot_oc}
\end{figure}

In this analysis, we used a new simple method 
for identification
of each starspot based on the rotational phase of local brightness 
minima in the light curve.
The light curve of the rotating star with a starspot group shows
a minimum brightness at the meridian passage of the spot during
one rotation period. The time separation of each photometric minimum
in the light curve is constant and equals to the rotation period.
In the case of the star with two spot groups, the light curve shows 
two local minima during the rotation period. 
The separation of the two local minima corresponds to the difference 
in longitudes between the two spots.
Moreover, if two spots are located at the different latitudes,
the time separation of two local minima, which 
correspond to the the difference in longitudes between each spot group,
changes in time due to the differential rotation of the star.
Figure \ref{spot_oc} shows some examples of the light curves of 
slowly-rotating stars (a, c, e, and g) and the time variations of the rotational 
phase of local brightness minima (b, d, f, and h).
All of the stars shown in figure \ref{spot_oc} are non-superflare stars.
The vertical axis of figure \ref{spot_oc} (b, d, f, and h) represents
the rotational phase of each local minimum in the light curve.
The phase of each minimum were calculated by using the rotation 
period of the star taken from \citet{McQuillan2014}.
If a series of local minima are caused by the same spot group, 
the rotational phase of the local minima lie on a line
on the phase-time plot until the spot group disappear.
Therefore, we can identify each starspot
and can estimate the lifetime of starspots by using our method.
The slope of the line corresponds to the difference in the 
apparent rotation period of the spot group due to the 
differential rotation. 
This indicates that we can also measure the differential rotation
in the stars from the standard deviation of the mean slope in the phase-time plots.
In figure \ref{spot_oc} (b),
we can clearly see two series of the local minima between BJD 2455300 and 2455450.
This result indicates
that there have been two major starspot groups on the surface of KIC 8282807
during that period.
If two or more spot groups are located at
the same longitude, the amplitude of photometric variations
would be large. As shown in figure \ref{spot_oc},
the two different spots were located at almost the same longitude
around BJD 2455500 and then the amplitude of photometric
variations became large.
This effect causes the overestimation of the area of 
starspots by using the equation (\ref{spotarea}).
However, if the latitudes of each spot are not the same,
the relative longitudes between each spots change in time
and we can estimate the area of each spot group from the light curve.
The light curves and the phase-time plots of these slowly-rotating stars indicate that
the number of major spot groups on the surface of the star at a given time is $1$-$3$. 

\begin{figure}
\begin{center}
\includegraphics[width=12cm]{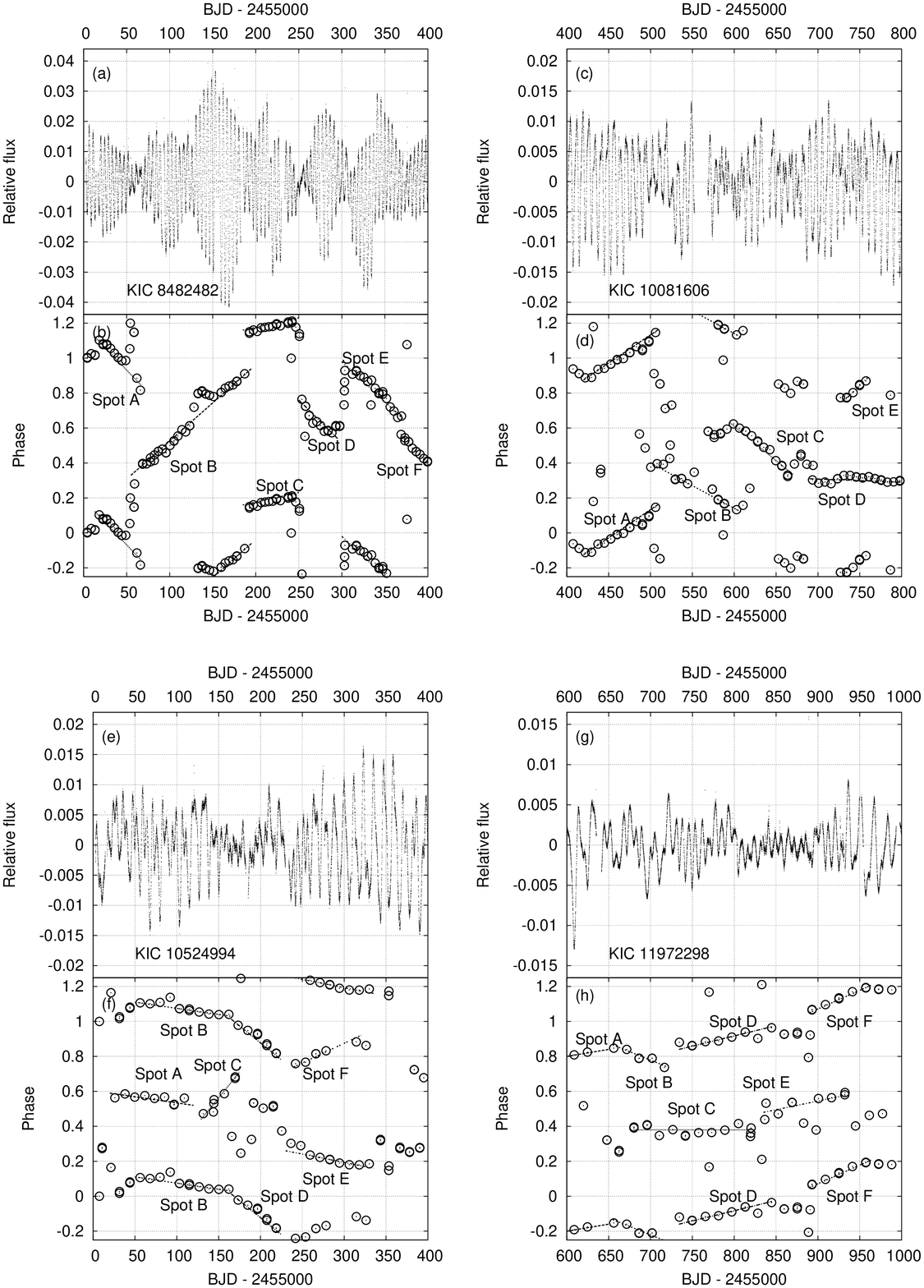}
\end{center}
\caption{
(a): Light curve of the solar-type superflare star KIC 8482482 ($P_{\rm rot}=4.47$ days).
(b): Time-variation of the rotational phase of the local photometric 
minima in the light curve of KIC 8482482.
(c) and (d): Same as panels (a) and (d), but for KIC 10081606 ($P_{\rm rot}=7.46$ days).
(e) and (f): Same as panels (a) and (d), but for KIC 10524994 ($P_{\rm rot}=11.9$ days).
(g) and (h): Same as panels (a) and (d), but for KIC 11972298 ($P_{\rm rot}=15.6$ days).
}
\label{spot_oc_short}
\end{figure}

We also performed the same analysis for the light curves of
the superflare stars with $P_{\rm rot}=4$-$15$ days.
Figure \ref{spot_oc_short} shows the light curves and phase-time plots
of KIC 8482482 ($P_{\rm rot}=4.47$ days),  KIC 10081606 ($P_{\rm rot}=7.46$ days), KIC 10524994 ($P_{\rm rot}=11.9$ days),
 and KIC 11972298 ($P_{\rm rot}=15.6$ days).
The superflare stars with the short- ($P_{\rm rot}<10$ days) and middle- ($P_{\rm rot}=10$-$20$ days)
rotation period also have a few major spots at the given time.
These results indicate
that the photometric amplitude can be used as a tracer of the spot area despite the rotation period
and flare activity of the star.

\begin{figure}
\begin{center}
\includegraphics[width=12cm]{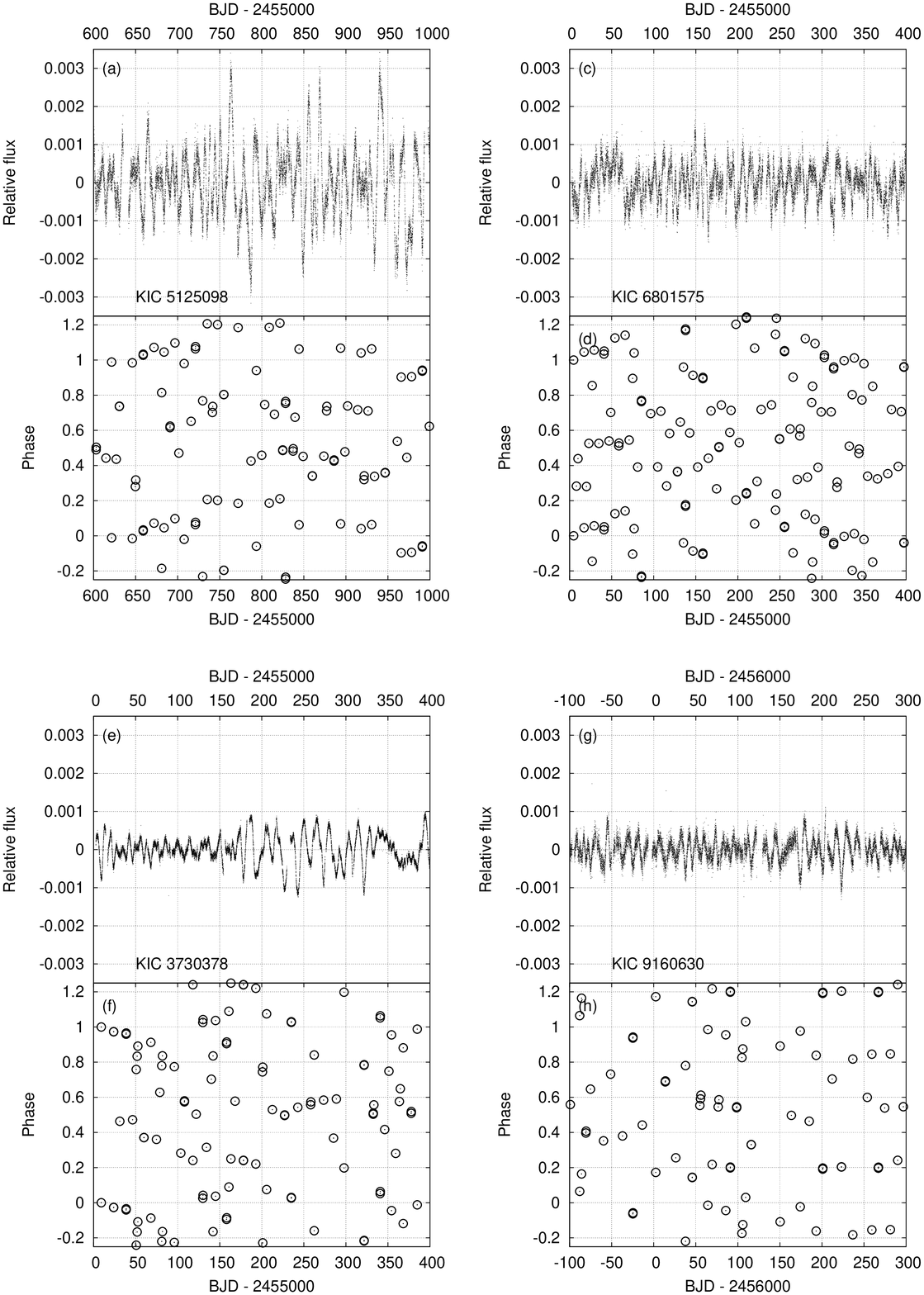}
\end{center}
\caption{
(a): Light curve of the solar-type star KIC 5125098 ($P_{\rm rot}=12.3$ days).
(b): Time-variation of the rotational phase of the local photometric 
minima in the light curve of KIC 5125098.
(c) and (d): Same as panels (a) and (d), but for KIC 6801575 ($P_{\rm rot}=11.9$ days).
(e) and (f): Same as panels (a) and (d), but for KIC 3730378 ($P_{\rm rot}=15.0$ days).
(g) and (h): Same as panels (a) and (d), but for KIC 9160630 ($P_{\rm rot}=22.0$ days).
}
\label{spot_oc_smallspots}
\end{figure}

The local photometric minima caused by the spot group 
lie on a line on the phase-time plot until the spot disappears.
Thus we can estimate the lifetime of each spot group
from the phase-time plots as shown in 
figure \ref{spot_oc} and \ref{spot_oc_short}.
These series of photometric local minima persist for 
a few hundred days (in the case of slowly-rotating stars) 
and $\sim 50$-$100$ days (rapidly-rotating stars), respectively.
These results indicate that the typical lifetime of large
starspot groups with the area of $\sim 10^{-2} A_{1/2\odot}$ on 
 solar-type stars is 
longer enough than the rotation period of the star.
In the case of slowly-rotating stars, the typical lifetime of
large spots is the order of $100$ days, which is much longer than
the lifetime of the typical sunspot groups ($<30$ days; e.g., \cite{Petrovay1997}).
In section 4.2, we assumed that the lifetime of spot groups
is longer than the rotation period. These results suggest
that the assumption on the spot lifetime is basically valid 
for the large starspots. 
In case of the starspots with the area of $\sim 10^{-3} A_{1/2\odot}$, 
the lifetime would be comparable to large sunspot groups.
We applied the same analysis to the light curves of the stars
showing low amplitude rotational modulations like the Sun.
Figure \ref{spot_oc_smallspots} represents the light curves and
phase-time plots of KIC 5125098 ($P_{\rm rot}=12.3$ days), KIC 6801575 ($11.9$ days), 
KIC 3730378 ($15.0$ days), and KIC 9160630 ($22.0$ days).
The area of starspots on these stars is estimated to be $\sim 2\times 10^{-3} A_{1/2\odot}$
from the equation (3).
We could not found any series of local minima which persist for 
$>100$ days from the phase-time plots of these stars.
The majority of series of local brightness minima disappear 
within $3$-$4$ stellar rotations, $\sim 50$ days, which is
comparable to the lifetime of sunspots with the area of 
$\sim 5\times 10^{-4} A_{1/2\odot}$ ($\sim 30$ days; \cite{Petrovay1997}).
These results suggest that it is possible to extend the size-lifetime 
relation of sunspots by \citet{Petrovay1997} up to the larger starspots
with the area of $\sim 10^{-2} A_{1/2\odot}$.



\section{Fraction of flare stars}
\begin{figure}
\begin{center}
\includegraphics{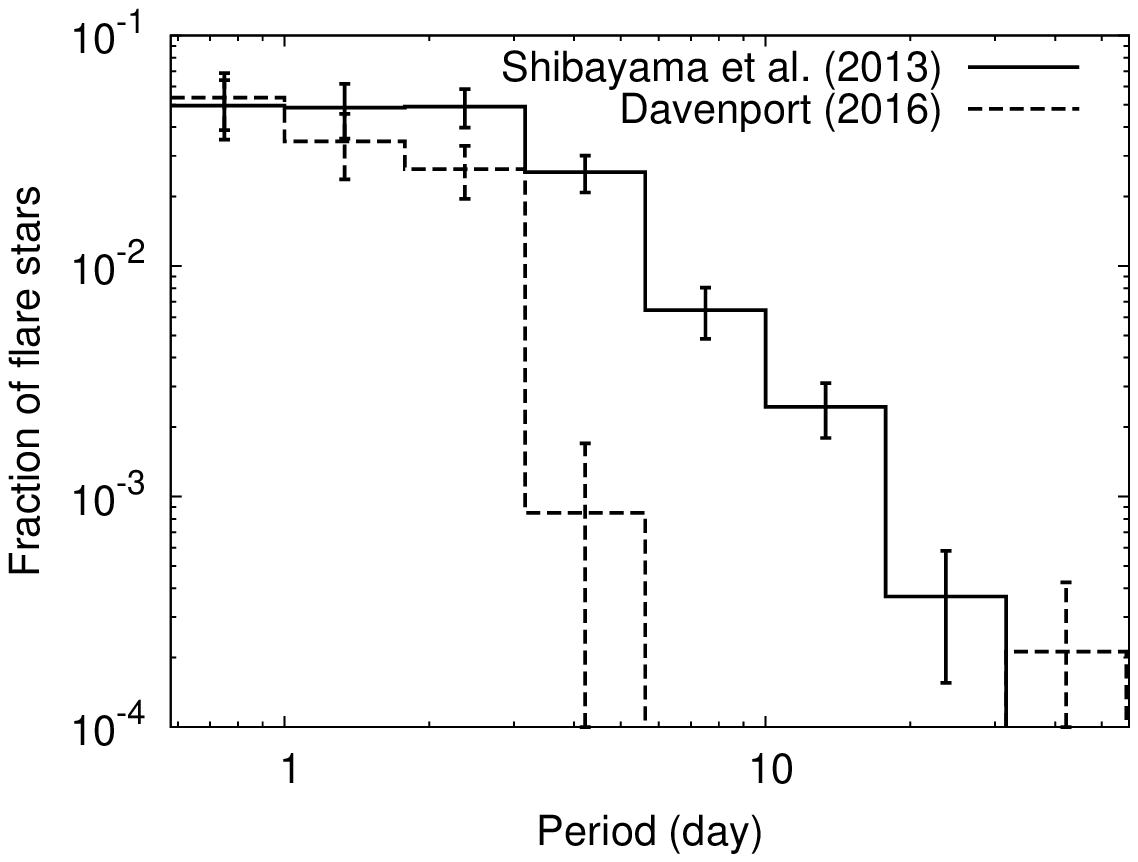}
\end{center}
\caption{
The fraction of flare stars as a function of the rotation period.
Solid line and dotted line represent the fraction of flare stars
derived from \citet{Shibayama2013} and that from \citet{Davenport2016}.
Since \citet{Davenport2016} selected only flare stars showing $>100$
flares during the Kepler observations, only the extremely active stars
would be listed in the list.
}
\label{Davenport2016_comp}
\end{figure}

As described in section 2, we used \citet{Shibayama2013} as the catalog of
superflare stars. Recently, \citet{Davenport2016} published the list of
flare stars in the Kepler field. 
Figure \ref{Davenport2016_comp} shows
the fraction of flare stars with $5600 \leq T_{\rm eff} < 6300$ K as a 
function of the rotation period derived from the list of flare stars
by \citet{Shibayama2013} and that by \citet{Davenport2016}.
In the short-period regime ($P_{\rm rot}<3$), the fraction of
flare stars derived from \citet{Davenport2016}
is comparable to that from \citet{Shibayama2013}.
On the other hand, in the period range of $P_{\rm rot}>3$ days,
the fraction of flare stars derived from \citet{Davenport2016}
is significantly lower than that from \citet{Shibayama2013}.
\citet{Davenport2016} selected only flare stars showing $>100$
flares during the Kepler observations (flare rate of $>26$ events per year).
This difference can be explained by the lack of 
moderately active superflare stars in the flare star list by \citet{Davenport2016}.
There are only a few flare stars with $P_{\rm rot}>3$ days in the list
by \citet{Davenport2016}. This suggests that the frequency of flares on
the flare stars with $P_{\rm rot}>3$ days are lower than that on the flare
stars with $P_{\rm rot}<3$ days.

%
%


\end{document}